\documentclass[a4paper,fleqn,usenatbib,letters]{mnras}

\usepackage{mathptmx}
\usepackage[T1]{fontenc}
\usepackage{ae,aecompl}

\usepackage{graphicx}	% Including figure files
\usepackage{amsmath}	% Advanced maths commands
\usepackage{amssymb}	% Extra maths symbols

\newcommand{\ha}{H$\alpha$}
\newcommand{\hb}{H$\beta$}

\newcommand{\hei}{He~{\sc i}}
\newcommand{\heii}{He~{\sc ii}}

\def\kms{\mbox{${\rm km}\:{\rm s}^{-1}\:$}}

\def\lesssim{\mathrel{\hbox{\rlap{\hbox{\lower4pt\hbox{$\sim$}}}\hbox{$<$}}}}
\let\la=\lesssim
\def\gtrsim{\mathrel{\hbox{\rlap{\hbox{\lower4pt\hbox{$\sim$}}}\hbox{$>$}}}}
\let\ga=\gtrsim

\title[The 2015 December mini-outburst of V404 Cygni]{Flares, wind and nebulae: the 2015 December mini-outburst of V404 Cygni} 
\author[Mu\~noz-Darias et al.]{T.~Mu\~noz-Darias$^{1,2}$, J.~Casares$^{1,2,3}$, D.~Mata S\'anchez$^{1,2}$, R.~P.~Fender$^{3}$,
\newauthor
M.~Armas Padilla$^{1,2}$, K.~Mooley$^{3,4}$, L.~Hardy$^{5}$, P.~A. Charles$^{3,6}$, G.~Ponti$^{7}$,
\newauthor
 S.~E.~Motta$^{3}$, V.~S.~Dhillon$^{1,5}$, P.~Gandhi$^{6}$, F.~Jim\'enez-Ibarra$^{1,2}$, T.~Butterley$^{8}$, 
\newauthor
S.~Carey$^{9}$, K.~J.~B.~Grainge$^{10}$, J.~Hickish$^{9}$, S.~ P.~Littlefair$^{5}$, Y.~C.~Perrott$^{9}$,  
\newauthor
N.~Razavi-Ghods$^{9}$, C.~Rumsey$^{9}$, A.~M.~M.~Scaife$^{10}$, P.~F.~Scott$^{9}$, 
\newauthor
D.~J.~Titterington$^{9}$, R.~ W.~ Wilson$^{8}$
\\
$^{1}$ Instituto de Astrof\'isica de Canarias, 38205 La Laguna, Tenerife, Spain \\
$^{2}$ Departamento de astrof\'isica, Univ. de La Laguna, E-38206 La Laguna, Tenerife, Spain \\
$^{3}$ Department of Physics, Astrophysics, University of Oxford, Keble Road, Oxford, OX1 3RH, UK \\
$^{4}$ Hintze Research Fellow \\
$^{5}$ Department of Physics \& Astronomy, University of Sheffield, Sheffield S3 7RH, UK \\
$^{6}$ School of Physics and Astronomy, University of Southampton, Highfield, Southampton SO17 1BJ, UK \\
$^{7}$ Max Planck Institute fur Extraterrestriche Physik, D-85748 Garching, Germany \\
$^{8}$ Department of Physics, University of Durham, South Road, Durham DH1 3LE, UK \\
$^{9}$ Astrophysics Group, Cavendish Laboratory, 19 J. J. Thomson Avenue, Cambridge CB3 0HE, UK \\
$^{10}$ Jodrell Bank Centre for Astrophysics, Alan Turing Building, Oxford Road, Manchester M13 9PL, UK \\
}
\date{Accepted XXX. Received YYY; in original form ZZZ}

\pubyear{2016}

\begin{document}
\label{firstpage}
\pagerange{\pageref{firstpage}--\pageref{lastpage}}
\maketitle

% Abstract of the paper
\begin{abstract}
After more than 26 years in quiescence, the black hole transient V404 Cyg went into a luminous outburst in June 2015, and additional activity was detected in late December of the same year. Here, we present an optical spectroscopic follow-up of the December mini-outburst, together with X-ray, optical and radio monitoring that spanned more than a month. Strong flares with gradually increasing intensity are detected in the three spectral ranges during the $\sim 10$ days following the Swift trigger. Our optical spectra reveal the presence of a fast outflowing wind, as implied by the detection of a P-Cyg profile (\hei --5876 \AA) with a terminal velocity of $\sim$ 2500 \kms. Nebular-like spectra -- with an \ha\ equivalent width of $\sim 500$ \AA\ -- are also observed. All these features are similar to those seen during the main June 2015 outburst. Thus, the fast optical wind simultaneous with the radio jet is most likely present in every V404 Cyg outburst. Finally, we report on the detection of a strong radio flare in late January 2016, when X-ray and optical monitoring had stopped due to Sun constraints.

\end{abstract}

% Select between one and six entries from the list of approved keywords.
% Don't make up new ones.

%%%%%%%%%%%%%%%%%%%%%%%%%%%%%%%%%%%%%%%%%%%%%%%%%%

%%%%%%%%%%%%%%%%% BODY OF PAPER %%%%%%%%%%%%%%%%%%

\section{Introduction}

\begin{figure*}
\begin{center}
\includegraphics[keepaspectratio,width=13.5cm]{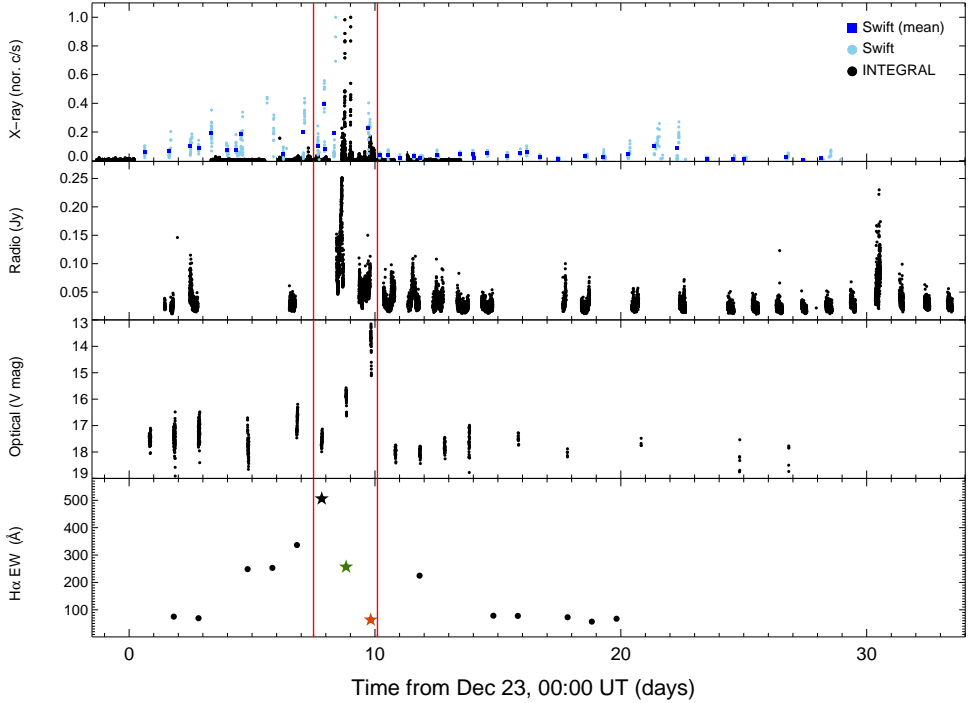}
\caption{From top to bottom X-ray (Swift and INTEGRAL), radio and optical lightcurves covering the December 2015 (to January 2016) mini-outburst of V404 Cyg. The bottom-most panel shows the \ha\ EW evolution derived from GTC and WHT spectra. The two vertical lines encompass the interval of highest activity, where conspicuous wind signatures are present in the spectra (marked as stars).}
\label{fig:lc}
\end{center}
\end{figure*}

\begin{figure}
\begin{center}
\includegraphics[keepaspectratio,width=8cm]{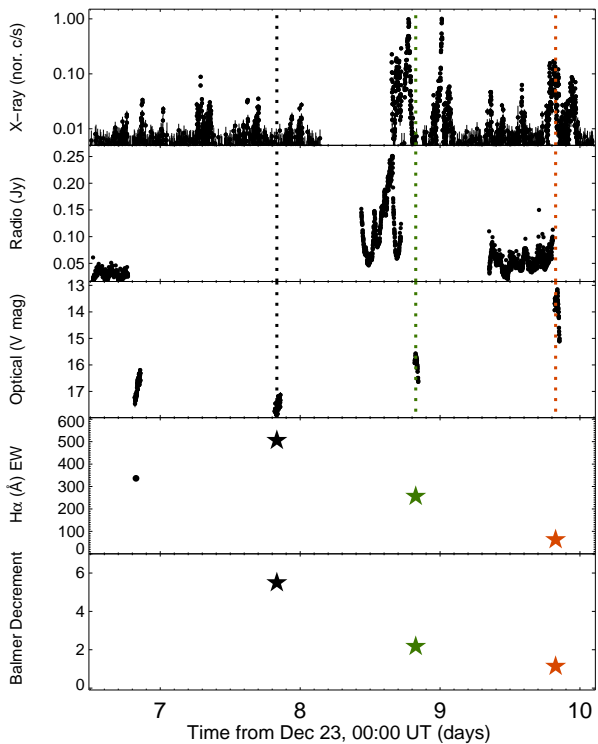}
\caption{Zoom of the interval from days 6.5 to 10, which encloses the peak of the outburst in all bands. The four top panels are the same as in Fig. \ref{fig:lc} (note logarithmic scale in the top-most), while the bottom panel represent the evolution of the Balmer decrement. Vertical dotted lines in the top three panels mark the times of the spectroscopic observations. Colour code is the same as in Fig. \ref{fig:lc}.}
\label{fig:zoom}
\end{center}
\end{figure}

\begin{figure*}
\begin{center}
\includegraphics[keepaspectratio,width=13.5cm]{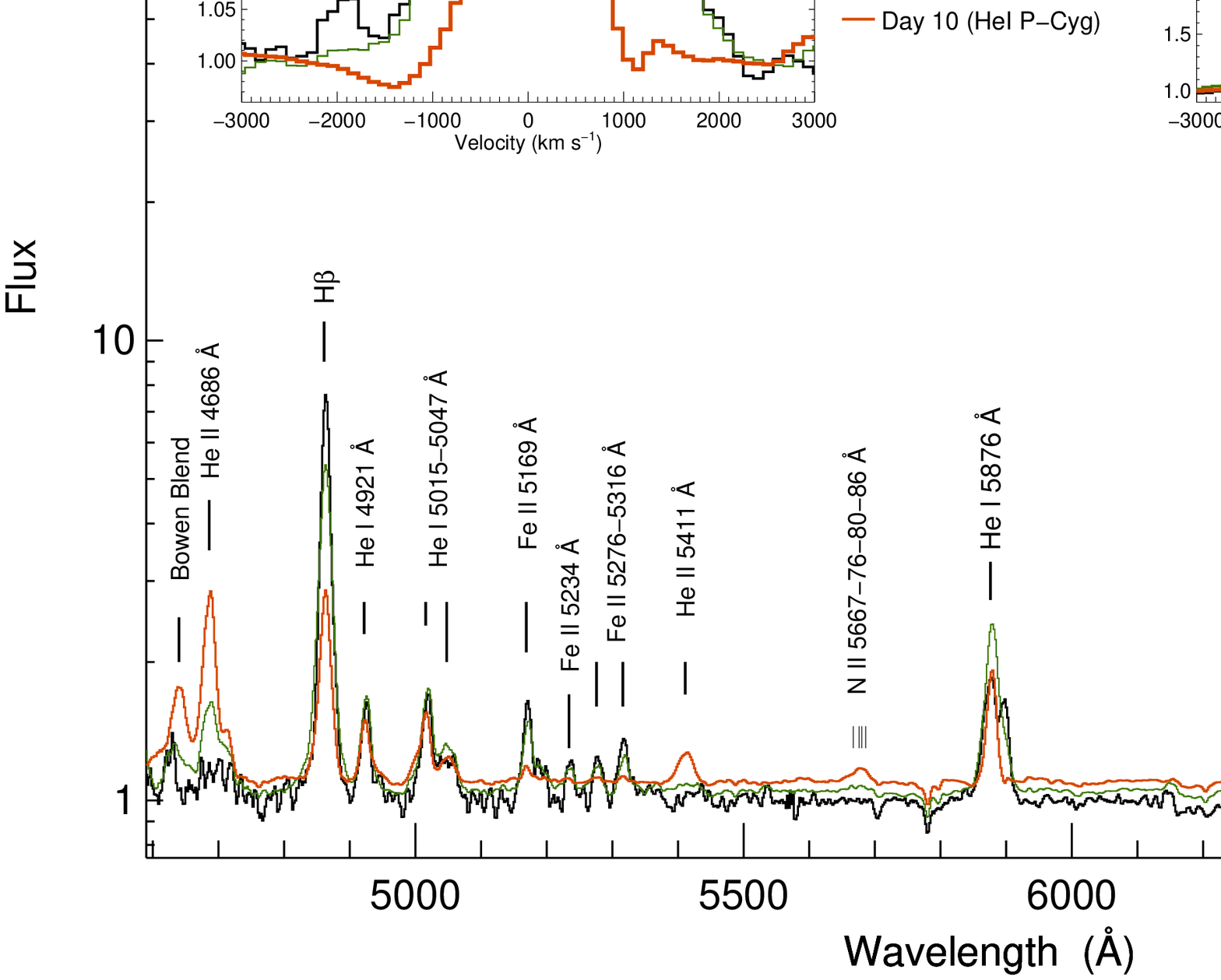}
\caption{Normalized spectra (on a logarithmic scale) corresponding to days 8--10. An offset of 0.05 and 0.1 has been added to the day-9 and day-10 spectra, respectively. Colour code is the same as for the stars in Fig. \ref{fig:lc}. The top-left inset shows a zoom of the \hei\ 5876 \AA\ region, where a P-Cyg profile is present on day 10. The top-right inset shows the \ha\ region. The emission line wings (day-8 and day-9) extend up to $\sim 2500$ \kms, a similar value to that of the blue absorption in the aforementioned P-Cyg profile.}
\label{fig:spectra}
\end{center}
\end{figure*}

\defcitealias{Munoz-Darias2016}{MD16}

V404 Cygni is a transient low mass X-ray binary harbouring a $\sim 9$ M$_{\odot}$ black hole and a K-type evolved donor in a 6.5 d orbital period, one of the longest of its class (\citealt{Casares1992,Casares2014,Corral-Santana2016}). It was discovered as an X-ray source by \textit{Ginga} during the peak of its 1989 outburst, when it displayed extreme X-ray behaviour in both the spectral and time domains (e.g. \citealt{Kitamoto1989,Zycki1999}). Photographic archives show that the system also went into outburst in 1938 and 1956  (\citealt{Richter1989,Wagner1991}). More recently, the \textit{Swift} mission detected renewed activity on Jun 15, 2015 \citep{Barthelmy2015}, which was the start of a luminous but brief outburst ($\sim 2$ weeks of strong activity) characterized by extreme phenomenology at high-energies \citep{Rodriguez2015,Siegert2016,Motta2016b}, optical \citep{Gandhi2016, Kimura2016} and radio bands \citep{Mooley2015}. Optical spectroscopy performed with the \textit{Gran Telescopio Canarias} (GTC)  10.4m telescope revealed the presence of an outflowing outer disc wind of low-ionisation/neutral material, detected as P-Cyg profiles in hydrogen (Balmer) and \hei\ emission lines \citep[][hereafter \citetalias{Munoz-Darias2016}]{Munoz-Darias2016}. The wind has a terminal velocity in the range 2000-3000 \kms\ and was observed simultaneously with the radio-jet, which is a distinctive feature of hard and intermediate black hole X-ray states \citep*{Fender2001a,Fender2016}. The wind is witnessed along the full active phase of the outburst, from X-ray fluxes as low as 10$^{-3}$ that of the outburst peak. Subsequently, an optically thin phase of the wind was observed following a sharp drop in flux by a factor $\sim 10^3$. This so-called \textit{nebular phase} lasted a few days and is characterized by broad emission lines -- equivalent width (EW) up to 2000 \AA\ (\ha) -- and large Balmer decrements ($\sim$6).  Six months after the June 2015 outburst (hereafter, June outburst), new activity was reported in V404 Cyg on Dec 23, 2015 \citep[MJD 57379, hereafter taken as time zero; ][]{Barthelmy2015b, Lipunov2015, Motta2015, Hardy2016}. However, the \textit{Fermi Gamma-ray Burst Monitor} \citep{Jenke2015} and INTEGRAL observations of the Cygnus region show that the source was active by Dec 21,  but it was not detected on Dec 13--15, 2015 \citep{Malyshev2015}.
In this letter, we present a 30-day X-ray, optical and radio monitoring campaign of V404 Cyg during Dec 2015--Jan 2016. We report on the observation of several intervals of flaring activity and the spectroscopic detection of an optical wind outflow similar to that observed during the June outburst.  
\section{Observations}
Optical photometric observations were conducted with the 0.5m robotic telescope, pt5m, situated at the \textit{Observatorio del Roque de los Muchachos} (ORM) on La Palma, Spain \citep{Hardy2015}. Photometry spanned over $\sim 25$ days (Fig \ref{fig:lc}) was obtained in the \textit{V} filter using 60 s, and later 120 s, exposures. The images were reduced in the standard manner using the ULTRACAM pipeline reduction software with variable-sized aperture photometry. Magnitudes were calibrated using the nearby comparison star USNO-B1.0 1238-0435227 ($V=12.82 \pm 0.05$; \citealt{Udalski1991}).

Optical spectroscopy was carried out over 14 different nights in the interval from 24 Dec 2015 to 11 Jan 2016 using the 10.4m GTC telescope (10 epochs) and the 4.2m William Herschel Telescope (WHT; 4 epochs) both located at the ORM. GTC observations were performed with the Optical System for Imaging and low/intermediate-Resolution Integrated Spectroscopy (OSIRIS) \citep{cepa2000} using grism R1000B, which covers the spectral range 3630 -- 7500 \AA\ with a resolving power of $R \sim 900$. The first three WHT observations (Dec 28, 29 and Jan 3) were performed using the  Intermediate dispersion Spectrograph and Imaging System (ISIS) with the R300B and R600R grisms in the blue and red arms, respectively. The WHT observation of Jan 9, 2016 was carried out using the auxiliary-port camera (ACAM), which provides $R \sim 450$ across the range 3500 -- 9400 \AA. Between 1 and 3 spectra were obtained every observing date, which were subsequently combined into one spectrum per epoch. 
Spectra were bias- and flat field-corrected using \textsc{iraf} routines. Due to the extremely limited object visibility, all the observations were taken at high airmass ($>2$) and with the slit positioned at the parallactic angle. Given these limitations, a reliable (absolute) flux calibration was not always possible. Indeed, a flux calibration was only performed for the highest signal-to-noise GTC spectra using the spectro-photometric standard G158-100. This is useful to constrain the Balmer decrement during the brightest epochs (marked by the solid, vertical lines in Fig. \ref{fig:lc}) . On the other hand, for every observation we computed the ionisation ratio ($I_{\mathrm ratio}$), defined as the flux ratio between the nearby \heii--4686 \AA\ and \hb\ emission lines, which is roughly independent of  flux calibration issues (see \citetalias{Munoz-Darias2016}).  

\subsection{X-ray and radio monitoring}
Following the same procedure as \citetalias{Munoz-Darias2016}, to which we refer the reader for further details on the X-ray analysis (methods section), we report on both INTEGRAL (20--250 keV; see \citealt{Kuulkers2016}) and Swift  (0.5 -- 10 keV) monitoring spanning 6 and 30 days, respectively (Fig. \ref{fig:lc}). Data from the Swift mission, including 41 different pointings, are presented at two different time resolutions, 100 s and $\sim 1$ ks (i.e. full observations). INTEGRAL observations started $\sim 2$ d after the Swift trigger and are shown in 64 s time steps. The peak count-rate in the 20--250 keV band was roughly a factor of 4 lower than that of the June outburst. Serendipitous INTEGRAL data obtained $\sim 2$ d prior to the Swift trigger are also included in Fig. \ref{fig:lc}  \citep[see][]{Malyshev2015}. 

Radio observations were performed with the Arcminute Microkelvin Imager Large Array (AMI-LA) radio telescope, operating as part of the University of Oxford 4-PI-SKY transients programme. These were made using the new digital correlator with 4096 channels across the 13--18 GHz frequency band. Observations started on 2015 December 24.69. We processed the data (RFI excision and calibration) with a fully-automated pipeline, AMIREDUCE \citep[e.g., ][]{Davies2009} and prepared images at a center frequency of 14 GHz.

Light curves were extracted in time steps of 40~s  and using six channels across the 5 GHz bandwidth via vector-averaging of the UV data.
V404 Cyg was observed for up to $\sim$ 12 hours per day for $\sim$30 days (Fig. \ref{fig:lc}). Many flaring episodes were detected during the campaign, reaching a maximum flux density of $\sim 0.25$ Jy at 16 GHz, a factor of 14 lower than the brightest flare during the June outburst (\citetalias{Munoz-Darias2016}; see Fender et al. [in prep]). 

\section{Results}

Our multi-wavelength coverage shows the presence of multiple flaring episodes during the $\sim 30$d following the Swift trigger (top three panels in Fig. \ref{fig:lc}). In X-rays, we detect increasing flaring activity up to day 10, being particularly intense on days 8--10 (see Fig. \ref{fig:zoom}). After this period, activity drops and only smaller flares on day $\sim 21$ are witnessed. Radio and optical observations are consistent with the X-ray picture, but a radio flare of comparable strength to that associated with the main X-ray episode (day 8) is detected on day 30, once optical and X-ray monitoring had already stopped due to Sun constraints.  Additional optical spectroscopy taken by this team once the object was again visible shows V404 Cyg in quiescence by late April 2016.

The system reached V$\sim13$ during the brightest flare (i.e. $\sim 2$ mags fainter than the peak of the June outburst), while fluxes only slightly above the quiescence level ($V\sim 18.5$) are measured outside the main activity period. Indeed, beyond day $\sim 11$, both optical and X-ray fluxes are mostly consistent with quiescence \citep[see also][]{Motta2016} and the spectra present double-peaked emission lines and companion star absorption features. The radio flux is always well above the $\sim 0.1$ mJy quiescence level, but we note that radio emission can be  observed for many days after a jet ejection episode (e.g. \citealt{Fender2007}).  Nevertheless, it is important to bear in mind that the system can show significant multiwavelength flaring activity even during true quiescence \citep{Zurita2003, Hynes2004b, Hynes2009}. \\      
The spectroscopic follow-up reveals dramatic changes in correspondence with the multi-wavelength variability of the source. The \ha\ EW is a reliable tracer -- independent of any flux calibration issue -- of the condition of the system. EWs > 200 \AA\ are observed between days 5--12, when the largest flares occur. During the June outburst, these large values are only measured either during strong wind (P-Cyg) phases at low $I_{\mathrm ratio}$ or along the final nebular phase (\citetalias{Munoz-Darias2016}).  On the other hand, the \ha\ EW is in the range 55--80 \AA\ outside the above interval, which is comparable with the lowest values observed during the June outburst.   Nevertheless, these are substantially larger than the $EW \sim$ 10--30 \AA\ observed in quiescence \citep{Casares2015}.\\
The large \ha\ EW observed on days 5, 6 and 7 (see bottom panel in Fig. \ref{fig:lc}) might suggest the presence of wind outflows at that stage of the outburst, probably in response to the increasing activity of the system. During days 8 to 10 we have solid evidence for this to be the case as both P-Cyg profiles and nebular-like spectra are witnessed (Figs. \ref{fig:lc}, \ref{fig:zoom} and \ref{fig:spectra}). 
On day 8, at low optical flux and following the detection of weak X-ray flares, a forest of Fe {\sc ii} emission lines become apparent in the spectrum. Also, the \ha\ EW reaches $\sim 500$ \AA\ and sits on wings extending up to $\sim 2500$ km s$^{-1}$ (Fig. \ref{fig:spectra}). This EW value exceeds any previous V404 Cyg measurement \citep[e.g.][]{Casares1991, Casares2015} with the remarkable exception of the nebular phase \textit{peak} of the June outburst (EW $\sim 2000$ \AA ; \citetalias{Munoz-Darias2016}). Indeed, it rivals the second largest value observed in June 2015. Similarly, on day 9, the \ha\ wings kept reaching the same velocity  ($\sim 2500$ km s$^{-1}$) even if the EW decreased by a factor of 2.  Finally, on day 10, we observed the system during a small X-ray/optical flare (Fig. \ref{fig:zoom}) and a weak (but obvious in our high signal-to-noise GTC spectra) P-Cyg profile is present in \hei --5876 \AA\ (Fig. \ref{fig:spectra}).  This shows a terminal velocity of $\sim 2500$ \kms and it is observed at $I_{\mathrm ratio}=1.2$, a value which is also associated with weak profiles during the June outburst. Besides this night (day-10), $I_{\mathrm ratio} \lesssim 0.2$ is measured owing to the weakness/absence of \heii-4686 (i.e. low ionisation state of the outer disc). We note that \hei --5876 \AA\  also shows the deepest P-Cyg profiles during the June outburst.\\
The combination of good observing conditions and source brightness allowed us to properly flux calibrate both the \hb\ and \ha\ regions during the day 8--10 period, and to determine the Balmer decrement, which decreases from $\sim 6$ on day 8 (nebular phase) to $\sim 2$ on day 10 (Fig. \ref{fig:zoom}). This behaviour is similar to that of the June outburst. Finally, we have computed the \ha\ line centroid velocity offset. Given the null systemic velocity and small velocity swing of the BH in V404 Cyg ($K_{\mathrm 1} \sim 12 $~\kms; \citealt{Casares1994}) any significant positive velocity offset could, in principle, be related to absorptions in the blue part of the profile and/or extra emission in the red wing, as observed during the June outburst (\citetalias{Munoz-Darias2016}). The three observations within the aforementioned interval (days 8--10) show offsets in the range 80--120~\kms , while $< 50$ \kms\ values are measured in the remaining observations.

\section{discussion}
\label{discussion}
We have presented X-ray, optical and radio light-curves covering the December 2015 outburst of V404 Cyg. With respect to the main (June) 2015 outburst, this accretion event was fainter by a factor of $\sim$ 4, 6 and 14 in the X-ray, optical and radio bands, respectively. However, we note that, in particular for the X-ray regime, absorption issues might be at work (e.g. \citealt{Motta2016b}) and a very detailed spectral study would be necessary to accurately measure the actual flux difference. Fainter sequels to bright outbursts are not unknown in black hole transients, probably the best cases being GRO~J0422+32 and XTE J1550-564, with significantly fainter events $\sim 300$ d after the main outburst \citep[e.g.][]{Castro-Tirado1997,Dunn2010}. Here, the mini-outburst occurred only $\sim 150$ d after the main but equally short ($\sim 3$ weeks) outburst. As noted in \citetalias{Munoz-Darias2016}, this time-scale might be consistent with the viscous time-scale to \emph{re-fill} the innermost part of the disc. Interestingly, although the optical and X-ray fluxes are consistent with quiescence only $\sim 10$ d after the Swift trigger, the system remained radio loud and displayed a second flare on day 30. It rivals in intensity the December event, showing that the latter was not unique. Indeed, since monitoring had stopped by late January due to Sun constraints, more activity in the following weeks cannot be ruled out.       

Besides the differing brightness, the evolution of the two 2015 episodes is remarkably similar, with multi-wavelength flux and flaring activity increasing on a time scale of $\sim$ 10 d from the Swift trigger, followed by a dramatic flux drop on a shorter interval (1--2 days). Similarly, the spectroscopic data shows that an optical wind, akin to that observed in June, was also present in some phases of the December event. This strongly suggests that this kind of outflow is not exclusive to bright outbursts [2015 (\citetalias{Munoz-Darias2016})  and 1989 \citep{Casares1991}], but is most likely a feature present in every V404 Cyg eruption, regardless of its brightness and duration. Likewise, the wind is, also in this case, simultaneous with the radio-jet, which is at odds with the usual evolution of the highly-ionised X-ray wind observed in several black hole and neutron star binaries (e.g. \citealt{Neilsen2009,Ponti2012,Ponti2014,Ponti2015}). However, after \citetalias{Munoz-Darias2016}, there is growing evidence of X-ray wind and radio-jet coexisting at least at some particular phases (\citealt{Homan2016, Kalemci2016}). It is probably too early to address the question of whether or not X-ray and optical winds are different signatures of the same process. In the particular case of V404 Cyg, we note that the optical wind and jet coexist during the whole June 2015 episode (i.e. from X-ray fluxes as low as $\sim 0.001$ times the outburst peak), and the X-ray wind is detected in the only observation performed by \textit{Chandra} \citep{King2015}.   
 
The two wind phases -- optically thick (P-Cyg profile) and optically thin (nebular) -- are most conspicuously observed during days 8 to 10 (vertical lines in Fig. \ref{fig:lc}).  In Fig. \ref{fig:zoom} it becomes evident that while the day-8 spectrum (nebular phase) was obtained during a quiet epoch following a series of flares, day-9 was taken in between two large X-ray flares at higher optical flux, and day-10 during the peak of an X-ray/optical/radio flare (Fig. \ref{fig:zoom}). Only in this latest case was a P-Cyg profile observed. We note that this is the only spectrum taken during a flare at $V\sim 13.5$, the remaining spectra corresponding to fainter epochs ($V>$15).  On the other hand, the June outburst nebular phase was only visible once the accretion dropped after $\sim 10$ days of activity, whilst here it is detected in the middle of the outburst before the strongest flares. This is most likely a result of the low flux level (roughly consistent with optical quiescence) reached by the system outside the main flares, in contrast to the higher activity always sustained in June. We conclude that the chronology of the different wind phases observed in December is affected by the particular short-term activity occurring at the time when the spectrum was taken. 

Our data only show wind signatures from day 5 at the earliest, when the flare intensity increases (e.g. Swift data in Fig. \ref{fig:lc}), whereas the June observations are consistent with the continuous presence of the wind. This raises the question of what is the low luminosity  boundary necessary to produce the wind, since our December data can be interpreted as wind being launched solely during the flares, after which the optically thin phase is observed. The thermal wind scenario \citep{Begelman1983} is consistent with the June 2015 observations (\citetalias{Munoz-Darias2016}), and can be extended -- given the similar observables -- to the data presented here. If this is the dominant wind-launching mechanism, the sustained higher activity implied by the lowest optical fluxes [$V\sim 17$ in June \citep{Kimura2016} vs. $V\sim 18.5$ in December] might be a determining factor. 

\section{Conclusions}
We have presented an optical spectroscopic follow-up, together with X-ray, optical and radio monitoring of the 2015 December mini-outburst of V404 Cyg. Despite this event being significantly fainter, the observed phenomenology resembles that of the June 2015 main outburst, including gradually brighter flares during the first 10 days of the outburst, followed by a much faster intensity drop. Likewise, we find solid evidence for the presence of an optical accretion disc wind in the form of P-Cyg profiles and nebular-like spectra.  New spectroscopic observations of other black hole transients and detailed modelling of the very rich June 2015 data should be able to shed more light on the properties and impact of this type of wind on accreting black holes.

\section*{Acknowledgements}

TMD, DMS and JC acknowledge support by the Spanish MINECO under grant AYA2013-42627. JC ackowledges support by the Leverhulme Trust (VP2-2015-04). We are thankful to the GTC and WHT teams that carried out the ToO observations. The AMI telescope gratefully acknowledges support from the European Research Council under grant ERC-2012- StG-307215 LODESTONE, the UK Science and Technology Facilities Council (STFC) and the University of Cambridge. KM research is supported by the Oxford Centre for Astrophysical Surveys, which is funded by the Hintze Family Charitable Foundation. GP acknowledges support by the German BMWI/DLR (FKZ 50 OR 1408 and FKZ 50 OR 1604) and the Max Planck Society. TB, RWW, VSD, SPL and PG acknowledge support by the UK STFC grants ST/J001236/1, ST/M001350/1 and ST/J003697/2.

%%%%%%%%%%%%%%%%%%%%%%%%%%%%%%%%%%%%%%%%%%%%%%%%%%

%%%%%%%%%%%%%%%%%%%% REFERENCES %%%%%%%%%%%%%%%%%%

% The best way to enter references is to use BibTeX:

\bibliographystyle{mnras}
\bibliography{/Users/tmd/Dropbox/Libreria} % if your bibtex file is called example.bib

\begin{thebibliography}{}
\makeatletter
\relax
\def\mn@urlcharsother{\let\do\@makeother \do\$\do\&\do\#\do\^\do\_\do\%\do\~}
\def\mn@doi{\begingroup\mn@urlcharsother \@ifnextchar [ {\mn@doi@}
  {\mn@doi@[]}}
\def\mn@doi@[#1]#2{\def\@tempa{#1}\ifx\@tempa\@empty \href
  {http://dx.doi.org/#2} {doi:#2}\else \href {http://dx.doi.org/#2} {#1}\fi
  \endgroup}
\def\mn@eprint#1#2{\mn@eprint@#1:#2::\@nil}
\def\mn@eprint@arXiv#1{\href {http://arxiv.org/abs/#1} {{\tt arXiv:#1}}}
\def\mn@eprint@dblp#1{\href {http://dblp.uni-trier.de/rec/bibtex/#1.xml}
  {dblp:#1}}
\def\mn@eprint@#1:#2:#3:#4\@nil{\def\@tempa {#1}\def\@tempb {#2}\def\@tempc
  {#3}\ifx \@tempc \@empty \let \@tempc \@tempb \let \@tempb \@tempa \fi \ifx
  \@tempb \@empty \def\@tempb {arXiv}\fi \@ifundefined
  {mn@eprint@\@tempb}{\@tempb:\@tempc}{\expandafter \expandafter \csname
  mn@eprint@\@tempb\endcsname \expandafter{\@tempc}}}

\bibitem[\protect\citeauthoryear{{Barthelmy}, {D'Ai}, {D'Avanzo}, {Krimm},
  {Lien}, {Marshall}, {Maselli}  \& {Siegel}}{{Barthelmy}
  et~al.}{2015a}]{Barthelmy2015}
{Barthelmy} S.~D.,  {D'Ai} A.,  {D'Avanzo} P.,  {Krimm} H.~A.,  {Lien} A.~Y.,
  {Marshall} F.~E.,  {Maselli} A.,   {Siegel} M.~H.,  2015a, GRB Coordinates
  Network, \href {http://adsabs.harvard.edu/abs/2015GCN..17929...1B} {17929}

\bibitem[\protect\citeauthoryear{{Barthelmy}, {Page}  \& {Palmer}}{{Barthelmy}
  et~al.}{2015b}]{Barthelmy2015b}
{Barthelmy} S.~D.,  {Page} K.~L.,   {Palmer} D.~M.,  2015b, GRB Coordinates
  Network, \href {http://adsabs.harvard.edu/abs/2015GCN..18716...1B} {18716}

\bibitem[\protect\citeauthoryear{{Begelman}, {McKee}  \& {Shields}}{{Begelman}
  et~al.}{1983}]{Begelman1983}
{Begelman} M.~C.,  {McKee} C.~F.,   {Shields} G.~A.,  1983, \mn@doi [\apj]
  {10.1086/161178}, \href {http://ads.nao.ac.jp/abs/1983ApJ...271...70B} {271,
  70}

\bibitem[\protect\citeauthoryear{{Casares}}{{Casares}}{2015}]{Casares2015}
{Casares} J.,  2015, \mn@doi [\apj] {10.1088/0004-637X/808/1/80}, \href
  {http://ads.nao.ac.jp/abs/2015ApJ...808...80C} {808, 80}

\bibitem[\protect\citeauthoryear{{Casares} \& {Charles}}{{Casares} \&
  {Charles}}{1994}]{Casares1994}
{Casares} J.,  {Charles} P.~A.,  1994, \mn@doi [\mnras]
  {10.1093/mnras/271.1.L5}, \href
  {http://adsabs.harvard.edu/abs/1994MNRAS.271L...5C} {271, L5}

\bibitem[\protect\citeauthoryear{{Casares} \& {Jonker}}{{Casares} \&
  {Jonker}}{2014}]{Casares2014}
{Casares} J.,  {Jonker} P.~G.,  2014, \mn@doi [\ssr]
  {10.1007/s11214-013-0030-6}, \href
  {http://ads.nao.ac.jp/abs/2014SSRv..183..223C} {183, 223}

\bibitem[\protect\citeauthoryear{{Casares}, {Charles}, {Jones}, {Rutten}  \&
  {Callanan}}{{Casares} et~al.}{1991}]{Casares1991}
{Casares} J.,  {Charles} P.~A.,  {Jones} D.~H.~P.,  {Rutten} R.~G.~M.,
  {Callanan} P.~J.,  1991, \mnras, \href
  {http://adsabs.harvard.edu/abs/1991MNRAS.250..712C} {250, 712}

\bibitem[\protect\citeauthoryear{{Casares}, {Charles}  \& {Naylor}}{{Casares}
  et~al.}{1992}]{Casares1992}
{Casares} J.,  {Charles} P.~A.,   {Naylor} T.,  1992, \mn@doi [\nat]
  {10.1038/355614a0}, \href {http://adsabs.harvard.edu/abs/1992Natur.355..614C}
  {355, 614}

\bibitem[\protect\citeauthoryear{{Castro-Tirado}, {Ortiz}  \&
  {Gallego}}{{Castro-Tirado} et~al.}{1997}]{Castro-Tirado1997}
{Castro-Tirado} A.~J.,  {Ortiz} J.~L.,   {Gallego} J.,  1997, \aap, \href
  {http://adsabs.harvard.edu/abs/1997A%26A...322..507C} {322, 507}

\bibitem[\protect\citeauthoryear{{Cepa} et~al.,}{{Cepa}
  et~al.}{2000}]{cepa2000}
{Cepa} J.,  et~al., 2000, in {Iye} M.,  {Moorwood} A.~F.,  eds,  SPIE
  Conference Series Vol. 4008, Optical and IR Telescope Instrumentation and
  Detectors. pp 623--631

\bibitem[\protect\citeauthoryear{{Corral-Santana}, {Casares},
  {Mu{\~n}oz-Darias}, {Bauer}, {Mart{\'{\i}}nez-Pais}  \&
  {Russell}}{{Corral-Santana} et~al.}{2016}]{Corral-Santana2016}
{Corral-Santana} J.~M.,  {Casares} J.,  {Mu{\~n}oz-Darias} T.,  {Bauer} F.~E.,
  {Mart{\'{\i}}nez-Pais} I.~G.,   {Russell} D.~M.,  2016, \mn@doi [\aap]
  {10.1051/0004-6361/201527130}, \href
  {http://ads.nao.ac.jp/abs/2016A%26A...587A..61C} {587, A61}

\bibitem[\protect\citeauthoryear{{Davies} et~al.,}{{Davies}
  et~al.}{2009}]{Davies2009}
{Davies} M.~L.,  et~al., 2009, \mn@doi [\mnras]
  {10.1111/j.1365-2966.2009.15518.x}, \href
  {http://ads.nao.ac.jp/abs/2009MNRAS.400..984D} {400, 984}

\bibitem[\protect\citeauthoryear{{Dunn}, {Fender}, {K{\"o}rding}, {Belloni}  \&
  {Cabanac}}{{Dunn} et~al.}{2010}]{Dunn2010}
{Dunn} R.~J.~H.,  {Fender} R.~P.,  {K{\"o}rding} E.~G.,  {Belloni} T.,
  {Cabanac} C.,  2010, \mn@doi [\mnras] {10.1111/j.1365-2966.2010.16114.x},
  \href {http://ads.nao.ac.jp/abs/2010MNRAS.403...61D} {403, 61}

\bibitem[\protect\citeauthoryear{{Fender}}{{Fender}}{2001}]{Fender2001a}
{Fender} R.~P.,  2001, \mn@doi [\mnras] {10.1046/j.1365-8711.2001.04080.x},
  \href {http://adsabs.harvard.edu/abs/2001MNRAS.322...31F} {322, 31}

\bibitem[\protect\citeauthoryear{{Fender} \& {Mu{\~n}oz-Darias}}{{Fender} \&
  {Mu{\~n}oz-Darias}}{2016}]{Fender2016}
{Fender} R.,  {Mu{\~n}oz-Darias} T.,  2016, in {Haardt} F.,  {et al.} eds,
  Lecture Notes in Physics Vol. 905, Berlin Springer Verlag. p.~65 (\mn@eprint
  {arXiv} {1505.03526}), \mn@doi{10.1007/978-3-319-19416-5_3}

\bibitem[\protect\citeauthoryear{{Fender}, {Dahlem}, {Homan}, {Corbel}, {Sault}
   \& {Belloni}}{{Fender} et~al.}{2007}]{Fender2007}
{Fender} R.~P.,  {Dahlem} M.,  {Homan} J.,  {Corbel} S.,  {Sault} R.,
  {Belloni} T.~M.,  2007, \mn@doi [\mnras] {10.1111/j.1745-3933.2007.00350.x},
  \href {http://ads.nao.ac.jp/abs/2007MNRAS.380L..25F} {380, L25}

\bibitem[\protect\citeauthoryear{{Gandhi} et~al.,}{{Gandhi}
  et~al.}{2016}]{Gandhi2016}
{Gandhi} P.,  et~al., 2016, \mn@doi [\mnras] {10.1093/mnras/stw571}, \href
  {http://ads.nao.ac.jp/abs/2016MNRAS.459..554G} {459, 554}

\bibitem[\protect\citeauthoryear{{Hardy}, {Butterley}, {Dhillon}, {Littlefair}
  \& {Wilson}}{{Hardy} et~al.}{2015}]{Hardy2015}
{Hardy} L.~K.,  {Butterley} T.,  {Dhillon} V.~S.,  {Littlefair} S.~P.,
  {Wilson} R.~W.,  2015, \mn@doi [\mnras] {10.1093/mnras/stv2279}, \href
  {http://ads.nao.ac.jp/abs/2015MNRAS.454.4316H} {454, 4316}

\bibitem[\protect\citeauthoryear{{Hardy}, {Gandhi}, {Dhillon}, {Littlefair},
  {Butterley}  \& {Wilson}}{{Hardy} et~al.}{2016}]{Hardy2016}
{Hardy} L.,  {Gandhi} P.,  {Dhillon} V.,  {Littlefair} S.,  {Butterley} T.,
  {Wilson} R.,  2016, The Astronomer's Telegram, \href
  {http://adsabs.harvard.edu/abs/2016ATel.8501....1H} {8501}

\bibitem[\protect\citeauthoryear{{Homan}, {Neilsen}, {Allen}, {Chakrabarty},
  {Fender}, {Fridriksson}, {Remillard}  \& {Schulz}}{{Homan}
  et~al.}{2016}]{Homan2016}
{Homan} J.,  {Neilsen} J.,  {Allen} J.~L.,  {Chakrabarty} D.,  {Fender} R.,
  {Fridriksson} J.~K.,  {Remillard} R.~A.,   {Schulz} N.,  2016, preprint,
  \href {http://adsabs.harvard.edu/abs/2016arXiv160607954H} {} (\mn@eprint
  {arXiv} {1606.07954})

\bibitem[\protect\citeauthoryear{{Hynes} et~al.,}{{Hynes}
  et~al.}{2004}]{Hynes2004b}
{Hynes} R.~I.,  et~al., 2004, \mn@doi [\apjl] {10.1086/424005}, \href
  {http://ads.nao.ac.jp/abs/2004ApJ...611L.125H} {611, L125}

\bibitem[\protect\citeauthoryear{{Hynes}, {Bradley}, {Rupen}, {Gallo},
  {Fender}, {Casares}  \& {Zurita}}{{Hynes} et~al.}{2009}]{Hynes2009}
{Hynes} R.~I.,  {Bradley} C.~K.,  {Rupen} M.,  {Gallo} E.,  {Fender} R.~P.,
  {Casares} J.,   {Zurita} C.,  2009, \mn@doi [\mnras]
  {10.1111/j.1365-2966.2009.15419.x}, \href
  {http://adsabs.harvard.edu/abs/2009MNRAS.399.2239H} {399, 2239}

\bibitem[\protect\citeauthoryear{{Jenke}, {WIlson-Hodge}  \&
  {Connaughton}}{{Jenke} et~al.}{2015}]{Jenke2015}
{Jenke} P.,  {WIlson-Hodge} C.~A.,   {Connaughton} V.,  2015, The Astronomer's
  Telegram, \href {http://adsabs.harvard.edu/abs/2015ATel.8457....1J} {8457}

\bibitem[\protect\citeauthoryear{{Kalemci}, {Begelman}, {Maccarone}, {Din{\c
  c}er}, {Russell}, {Bailyn}  \& {Tomsick}}{{Kalemci}
  et~al.}{2016}]{Kalemci2016}
{Kalemci} E.,  {Begelman} M.~C.,  {Maccarone} T.~J.,  {Din{\c c}er} T.,
  {Russell} T.~D.,  {Bailyn} C.,   {Tomsick} J.~A.,  2016, \mn@doi [\mnras]
  {10.1093/mnras/stw2002}, \href
  {http://adsabs.harvard.edu/abs/2016MNRAS.463..615K} {463, 615}

\bibitem[\protect\citeauthoryear{{Kimura} et~al.,}{{Kimura}
  et~al.}{2016}]{Kimura2016}
{Kimura} M.,  et~al., 2016, \mn@doi [\nat] {10.1038/nature16452}, \href
  {http://ads.nao.ac.jp/abs/2016Natur.529...54K} {529, 54}

\bibitem[\protect\citeauthoryear{{King}, {Miller}, {Raymond}, {Reynolds}  \&
  {Morningstar}}{{King} et~al.}{2015}]{King2015}
{King} A.~L.,  {Miller} J.~M.,  {Raymond} J.,  {Reynolds} M.~T.,
  {Morningstar} W.,  2015, \mn@doi [\apjl] {10.1088/2041-8205/813/2/L37}, \href
  {http://ads.nao.ac.jp/abs/2015ApJ...813L..37K} {813, L37}

\bibitem[\protect\citeauthoryear{{Kitamoto}, {Tsunemi}, {Miyamoto}, {Yamashita}
   \& {Mizobuchi}}{{Kitamoto} et~al.}{1989}]{Kitamoto1989}
{Kitamoto} S.,  {Tsunemi} H.,  {Miyamoto} S.,  {Yamashita} K.,   {Mizobuchi}
  S.,  1989, \mn@doi [\nat] {10.1038/342518a0}, \href
  {http://ads.nao.ac.jp/abs/1989Natur.342..518K} {342, 518}

\bibitem[\protect\citeauthoryear{{Kuulkers}, {amp}  \& {Ferrigno}}{{Kuulkers}
  et~al.}{2016}]{Kuulkers2016}
{Kuulkers} E.,  {amp}  {Ferrigno} C.,  2016, The Astronomer's Telegram, \href
  {http://adsabs.harvard.edu/abs/2016ATel.8512....1K} {8512}

\bibitem[\protect\citeauthoryear{{Lipunov} et~al.,}{{Lipunov}
  et~al.}{2015}]{Lipunov2015}
{Lipunov} V.,  et~al., 2015, The Astronomer's Telegram, \href
  {http://adsabs.harvard.edu/abs/2015ATel.8453....1L} {8453}

\bibitem[\protect\citeauthoryear{{Malyshev}, {Savchenko}, {Ferrigno}, {Bozzo}
  \& {Kuulkers}}{{Malyshev} et~al.}{2015}]{Malyshev2015}
{Malyshev} D.,  {Savchenko} V.,  {Ferrigno} C.,  {Bozzo} E.,   {Kuulkers} E.,
  2015, The Astronomer's Telegram, \href
  {http://adsabs.harvard.edu/abs/2015ATel.8458....1M} {8458}

\bibitem[\protect\citeauthoryear{{Mooley}, {Fender}, {Anderson}, {Staley},
  {Kuulkers}  \& {Rumsey}}{{Mooley} et~al.}{2015}]{Mooley2015}
{Mooley} K.,  {Fender} R.,  {Anderson} G.,  {Staley} T.,  {Kuulkers} E.,
  {Rumsey} C.,  2015, The Astronomer's Telegram, \href
  {http://adsabs.harvard.edu/abs/2015ATel.7658....1M} {7658}

\bibitem[\protect\citeauthoryear{{Motta} et~al.,}{{Motta}
  et~al.}{2015}]{Motta2015}
{Motta} S.~E.,  et~al., 2015, The Astronomer's Telegram, \href
  {http://adsabs.harvard.edu/abs/2015ATel.8462....1M} {8462}

\bibitem[\protect\citeauthoryear{{Motta}, {Kajava},
  {S{\'a}nchez-Fern{\'a}ndez}, {Giustini}  \& {Kuulkers}}{{Motta}
  et~al.}{2016a}]{Motta2016b}
{Motta} S.~E.,  {Kajava} J.~J.~E.,  {S{\'a}nchez-Fern{\'a}ndez} C.,  {Giustini}
  M.,   {Kuulkers} E.,  2016a, preprint, \href
  {http://adsabs.harvard.edu/abs/2016arXiv160702255M} {} (\mn@eprint {arXiv}
  {1607.02255})

\bibitem[\protect\citeauthoryear{{Motta} et~al.,}{{Motta}
  et~al.}{2016b}]{Motta2016}
{Motta} S.~E.,  et~al., 2016b, The Astronomer's Telegram, \href
  {http://adsabs.harvard.edu/abs/2016ATel.8510....1M} {8510}

\bibitem[\protect\citeauthoryear{{Mu{\~n}oz-Darias} et~al.,}{{Mu{\~n}oz-Darias}
  et~al.}{2016}]{Munoz-Darias2016}
{Mu{\~n}oz-Darias} T.,  et~al., 2016, \mn@doi [\nat] {10.1038/nature17446},
  \href {http://ads.nao.ac.jp/abs/2016Natur.534...75M} {534, 75}

\bibitem[\protect\citeauthoryear{{Neilsen} \& {Lee}}{{Neilsen} \&
  {Lee}}{2009}]{Neilsen2009}
{Neilsen} J.,  {Lee} J.~C.,  2009, \mn@doi [\nat] {10.1038/nature07680}, \href
  {http://ads.nao.ac.jp/abs/2009Natur.458..481N} {458, 481}

\bibitem[\protect\citeauthoryear{{Ponti}, {Fender}, {Begelman}, {Dunn},
  {Neilsen}  \& {Coriat}}{{Ponti} et~al.}{2012}]{Ponti2012}
{Ponti} G.,  {Fender} R.~P.,  {Begelman} M.~C.,  {Dunn} R.~J.~H.,  {Neilsen}
  J.,   {Coriat} M.,  2012, \mn@doi [\mnras]
  {10.1111/j.1745-3933.2012.01224.x}, \href
  {http://ads.nao.ac.jp/abs/2012MNRAS.422L..11P} {422, L11}

\bibitem[\protect\citeauthoryear{{Ponti}, {Mu{\~n}oz-Darias}  \&
  {Fender}}{{Ponti} et~al.}{2014}]{Ponti2014}
{Ponti} G.,  {Mu{\~n}oz-Darias} T.,   {Fender} R.~P.,  2014, \mn@doi [\mnras]
  {10.1093/mnras/stu1742}, \href {http://ads.nao.ac.jp/abs/2014MNRAS.444.1829P}
  {444, 1829}

\bibitem[\protect\citeauthoryear{{Ponti} et~al.,}{{Ponti}
  et~al.}{2015}]{Ponti2015}
{Ponti} G.,  et~al., 2015, \mn@doi [\mnras] {10.1093/mnras/stu1853}, \href
  {http://adsabs.harvard.edu/abs/2015MNRAS.446.1536P} {446, 1536}

\bibitem[\protect\citeauthoryear{{Richter}}{{Richter}}{1989}]{Richter1989}
{Richter} G.~A.,  1989, Information Bulletin on Variable Stars, \href
  {http://adsabs.harvard.edu/abs/1989IBVS.3362....1R} {3362}

\bibitem[\protect\citeauthoryear{{Rodriguez} et~al.,}{{Rodriguez}
  et~al.}{2015}]{Rodriguez2015}
{Rodriguez} J.,  et~al., 2015, \mn@doi [\aap] {10.1051/0004-6361/201527043},
  \href {http://ads.nao.ac.jp/abs/2015A%26A...581L...9R} {581, L9}

\bibitem[\protect\citeauthoryear{{Siegert} et~al.,}{{Siegert}
  et~al.}{2016}]{Siegert2016}
{Siegert} T.,  et~al., 2016, \mn@doi [\nat] {10.1038/nature16978}, \href
  {http://ads.nao.ac.jp/abs/2016Natur.531..341S} {531, 341}

\bibitem[\protect\citeauthoryear{{Udalski} \& {Kaluzny}}{{Udalski} \&
  {Kaluzny}}{1991}]{Udalski1991}
{Udalski} A.,  {Kaluzny} J.,  1991, \mn@doi [\pasp] {10.1086/132808}, \href
  {http://ads.nao.ac.jp/abs/1991PASP..103..198U} {103, 198}

\bibitem[\protect\citeauthoryear{{Wagner}, {Bertram}, {Starrfield}, {Howell},
  {Kreidl}, {Bus}, {Cassatella}  \& {Fried}}{{Wagner}
  et~al.}{1991}]{Wagner1991}
{Wagner} R.~M.,  {Bertram} R.,  {Starrfield} S.~G.,  {Howell} S.~B.,  {Kreidl}
  T.~J.,  {Bus} S.~J.,  {Cassatella} A.,   {Fried} R.,  1991, \mn@doi [\apj]
  {10.1086/170429}, \href {http://ads.nao.ac.jp/abs/1991ApJ...378..293W} {378,
  293}

\bibitem[\protect\citeauthoryear{{Zurita}, {Casares}  \& {Shahbaz}}{{Zurita}
  et~al.}{2003}]{Zurita2003}
{Zurita} C.,  {Casares} J.,   {Shahbaz} T.,  2003, \mn@doi [\apj]
  {10.1086/344534}, \href {http://ads.nao.ac.jp/abs/2003ApJ...582..369Z} {582,
  369}

\bibitem[\protect\citeauthoryear{{{\.Z}ycki}, {Done}  \& {Smith}}{{{\.Z}ycki}
  et~al.}{1999}]{Zycki1999}
{{\.Z}ycki} P.~T.,  {Done} C.,   {Smith} D.~A.,  1999, \mn@doi [\mnras]
  {10.1046/j.1365-8711.1999.02885.x}, \href
  {http://ads.nao.ac.jp/abs/1999MNRAS.309..561Z} {309, 561}

\makeatother
\end{thebibliography}

%%%%%%%%%%%%%%%%%%%%%%%%%%%%%%%%%%%%%%%%%%%%%%%%%%

%%%%%%%%%%%%%%%%% APPENDICES %%%%%%%%%%%%%%%%%%%%%

%%%%%%%%%%%%%%%%%%%%%%%%%%%%%%%%%%%%%%%%%%%%%%%%%%

% Don't change these lines
\bsp	% typesetting comment
\label{lastpage}
\end{document}